\documentclass[aps,prl,10pt,floatfix,longbibliography,twocolumn]{revtex4-1}

\usepackage{graphicx}
\usepackage{amsfonts}
\usepackage{amsmath}
\usepackage{amssymb}

\usepackage[T1]{fontenc} % Modern font encoding
\usepackage{float}       % For creating charts, graphs and schemes

\begin{document}

\title{Interaction as stochastic noise}
\author{Roberto D'Agosta}
\affiliation{Nano-bio Spectroscopy Group, Departamento de Fisica de Materiales, UPV/EHU, San Sebastian, 20018 Spain, and IKERBASQUE, Basque Foundation for Science, E-48013, Bilbao, Spain}
\email{roberto.dagosta@ehu.es}

\date{\today}
\maketitle

{\bf Interaction is so ubiquitous that imaging a world free from it is a
difficult fantasy exercise. At the same time, in understanding any complex
physical system, our ability of accounting for the mutual interaction of its
constituents is often insufficient when not the restraining factor. Many
strategies have been devised to control particle-particle interaction and
explore the diverse regimes, from weak to strong interaction. Beautiful
examples of these achievements are the experiments on Bose condensates
\cite{Pollack2009,Dalfovo1999,Nguyen2014}, or the recent experiments on the
dynamics of spin chains~\cite{Jurcevic2014,Richerme2014}. Here I introduce
another possibility, namely replacing the particle-particle interaction with an
external stochastic field, and once again reducing the dynamics of a many-body
system to the dynamics of single-particle systems. The theory is exact, in the
sense that no approximations are introduced in decoupling the many-body system
in its non-interacting sub-parts. Moreover, the equations of motion are linear,
and no unknown external potential is inserted.}

The idea of replacing the many-body system under investigation with a
non-interacting doppelganger is not new. From a theoretical point of view, a
starting idea has been to treat the interaction as an external perturbation.
Interaction ``dresses'' the particles and new fundamental particles appear for
the description of a physical phenomenon. This is the tenet of the Landau's
theory of the Fermi gas mapping a strongly interacting electron gas into a
system of weakly interacting quasi-particles
\cite{Mattuck1992,Giulianivignale}. More modern approaches replace the
particle-particle interaction with an external effective potential, e.g., the
Thomas-Fermi's theory and the Hartree-Fock approximation \cite{Bransden1995}
which have all converged now somewhat in the Density Functional
Theory~\cite{Hohenberg1964,Kohn1965,Giulianivignale}. The price to pay for this
huge simplification is the inclusion of a unknown non-linear potential in the
dynamics of the fictitious non-interacting
system~\cite{Giulianivignale,Dreizler1990}. Density functional theory has been
instrumental in understanding many physical, chemical, and biological phenomena
at the nano-scale and in augmenting the theoretical prediction potential.

With hindsight the results I will present in the following are not completely
surprising: for example when dealing with magnetic systems, a common
approximation consists in replacing the dynamics of spin operators with the
dynamics of their quantum averages. Often, these averages have a random
behavior since they mostly consist of the superposition of a static magnetic
moment and small dynamical fluctuations.

In this Letter, I will show that this analogy is even more stringent and can be
made exact for the case in which particles interact via a potential that
depends on one operator of particle $i$ multiplied by an operator of particle
$j$. Indeed, the dynamics of such a system is exactly equivalent to the
dynamics of a system of non-interacting particles in the presence of known
stochastic potentials. The dynamics of the many-body system is then recovered
by constructing the proper wave-function or density matrix and averaging the
results over the stochastic fields. This study finds direct application in the
dynamics of arbitrary spin-chains and gives immediate access to the exact
high-order correlation functions that recently have been probed experimentally
\cite{Jurcevic2014,Richerme2014}. Moreover, the results presented here open up
the possibility of investigating the dynamics of large systems. It is well
known indeed that the numerical solution of the equation of motion of the
many-body density matrix, $\hat \rho$, is limited by its intrinsic large
dimensionality which scales at least quadratically with the number of states
one needs to consider, compared with the linear scaling of the wave-function.
Our result shows that for certain cases, one can bring the exact dynamics down
to the evaluation of $N$ relatively small density matrices, thus regaining a
linear scaling. To give an estimate of the problem with the interacting system,
with a chain of $N$ spin $1/2$, the exact many-body states belongs to a space
of dimensionality $2^N$, and therefore the density matrix has dimension
$2^N\times 2^N$. It should be then apparent that we can investigate numerically
small chains, routinely up to $N\simeq10-15$, after which the computer memory
requirements will be prevailing. Our result brings down this requirement to
store up to $N$ $2\times 2$ matrices.

Let us begin with considering a many-body quantum system, whose dynamics is
determined by the many-body Hamiltonian 
\begin{equation} \hat H=\sum_i^N \hat
h_i+\sum_{i<j}^N\lambda_{i,j}\hat x_i \otimes \hat x_j, 
\label{totH}
\end{equation} 
where $h_i$ is a single particle Hamiltonian and $\hat x_i$ is
some operator acting on the particle $i$, $\lambda_{i,j}$ the interaction
constant between particle $i$ and particle $j$, and $N$ the total particle
number. By Newton's third law, the interaction constant is symmetric, i.e,
$\lambda_{i,j}=\lambda_{j,i}$. The many-body density matrix of the system
evolves according to the von Neumann's equation 
\begin{equation} 
	\hbar\partial_t\hat\rho(t)=-i\left[\hat H(t),\hat \rho(t)\right]. 
\label{von_neumann} 
\end{equation} 
For simplicity, let us
assume that the initial density matrix is the direct product of single particle
density matrices 
\begin{equation}
\hat \rho(t=0)=\bigotimes_{i=1,N}\hat
\rho_i(t=0), 
\label{inital_state} 
\end{equation} 
then we can find a set of
$N^2$ independent white complex noises $\omega_{i,j}(t)$ for which the {\emph
single particle} density matrix $\hat \rho_i$ evolves according
to~\footnote{Eq.~(\ref{effective_dynamics}) is not unique. Indeed, a family of
equivalent equations can be easily obtained by shifting the weight of the
interaction between the terms in the round bracket in the right hand side of
that equation. The form chosen here keeps a balanced weight between the two
terms. None of the physical results depend on the details of this choice.}
\begin{equation} 
	\begin{split}
\hbar d\hat\rho_i(t)=&-i\left[h_i,\hat\rho_i(t)\right]dt\\
&+\sum_{j=1}^N\frac{\sqrt{\hbar\lambda_{i,j}}}{2}\left(\left[\hat x_i,\hat
\rho_i\right]d\omega_{i,j}-i\left\{\hat x_i,\hat \rho_i
\right\}d\omega^*_{j,i}\right)
 \label{effective_dynamics} 
\end{split}
\end{equation} 
and the exact total density matrix is given by 
\begin{equation}
\hat \rho(t)=\overline{\bigotimes_{i=1,N}\hat\rho_i(t)} 
\label{exactrho}
\end{equation} 
where the $\overline\cdots$ denotes the average over all the
white noises. In Eqs.~(\ref{von_neumann}) and (\ref{effective_dynamics}), $[\hat A,\hat B ]=\hat A \hat
B-\hat B\hat A$ and $\{\hat A,\hat B\}=\hat A\hat B+\hat B\hat A$ are the
standard commutator and anti-commutator of any two operators, respectively. The
white noises $\omega_{i,j}$ are complex Wiener processes that satisfy \begin{equation}
\overline{d\omega_{i,j}^*d\omega_{k,p}}=2\delta_{i,k}\delta_{j,p}dt,
\label{white_noise} \end{equation} 
where $\delta_{i,k}=1$ if $i=k$ or vanishes otherwise
\cite{Gardiner1983,Gardiner2000,Breuer2002,Higham2001}. We prove Eq.~(\ref{exactrho}) starting from Eqs.~(\ref{effective_dynamics}) in the Methods section. 

Let me now discuss two important points. First, it may appear that the initial
correlation between the particles is lost in this theory. This is not the case.
Indeed, the following discussion can be generalized to the case in which the
initial condition is given by $ \hat \rho(0)=\bigotimes_{i=1,N} \sum_k
\gamma_k\hat \rho_i^k(0), $ where ${\gamma_k}$ is a set of coefficients such
that $\sum_k\gamma_k=1$, to ensure that if $\mathrm{tr}\hat \rho_i^k(0)=1$ for
any $i$, then $\mathrm{tr}\hat \rho(0)=1$. The total density matrix in this
case, owing to the linearity of the equation of motion, will then be given by $
\hat \rho(t)=\sum_k\gamma_k\overline{\bigotimes_{i=1,N}\hat\rho^k_i(t)}, $
where each $\hat \rho^k_i$ evolves according to Eq.~(\ref{effective_dynamics})
with initial condition $\hat \rho^k_i(0)$. For simplicity, in the following I
will maintain that Eq.~(\ref{inital_state}) is satisfied by the initial density
matrices. The second point is the form of the particle-particle interaction: if
any operator of particle $i$ commutes, or anti-commutes, with any operator of
particle $j$ then one can always expand any two-particle interaction as a
series of products of single particle operators like in Eq.~(\ref{totH}). The
theory we are putting forward then is useful for those case in which the
particle-particle interaction is written as a finite sum of products of pairs
of single-particle operators. Finally, it should be clear that even if the
initial density matrix $\hat\rho_i(0)$ describes a real particle, its time
evolution $\hat \rho_i(t)$ cannot be associated with the dynamics of a real
particle. This is easily seen by the fact that Eq.~(\ref{effective_dynamics})
does not preserve either the positivity or the unitarity of $\hat\rho_i(t)$,
even if we assume $\hat \rho_i(0)$ is a definite positive matrix of unitary
trace. We will discuss later on how to reduce the many-body density matrix to a
proper single-particle (reduced) density matrix that can be used to investigate
the single-particle properties of the system.

The exact dynamics of the many-body density matrix usually contains a redundant
amount of information. For practical purposes, it is usually more convenient to
trace out some of the degrees of freedom and obtain the expectation
value of single- or two-particle operators. Within this theory, this procedure
emerges naturally from the definition of the operator $\hat \rho_i$. We have for example for the time evolution of the subsystem $j$,
\begin{equation}
	\hat \rho_j^R(t)=\overline{\hat\rho_j(t)\prod_{i\not= j}\mathrm{tr}\hat \rho_i(t) },
	\label{reduced_dynamics}
\end{equation}
where $\mathrm{tr}$ indicates the trace operation on the matrix $\hat \rho_i$. 
Notice that in general $\overline{\mathrm{tr}\hat\rho_i(t)}$ is a function of
time, and we cannot expect that it equals 1 at all times. On the other hand,
we expect that $\mathrm{tr}\hat\rho_j^R(t)=\mathrm{tr}\hat\rho(t)=1$ at all times, so
that $\hat\rho_j^R$ does define a proper density matrix for the subsystem
$j$~\footnote{We leave out the question if $\hat\rho_j^R(t)$ is also definitive
positive. This point requires further investigation.}

As a first example of application we consider the case of two interacting spins. We assume there is not any external magnetic field. The Hamiltonian for this simple system is
$
	\hat H=\lambda \sigma_1^z\otimes\sigma_2^z
$
where $\sigma_i^z$ is the $2\times 2$ third Pauli matrix. For simplicity and without any loss of generality we can set $\lambda=\hbar$. According to Eq.~(\ref{effective_dynamics}) we need to solve the two independent stochastic master equations
\begin{equation}
	\begin{split}
	d\hat\rho_1=\frac{1}{2}\left[\sigma^z\cdot\hat\rho_1(d\omega_{12}-id\omega^*_{21})-
	\hat\rho_1\cdot\sigma^z(id\omega^*_{21}+d\omega_{12})\right],\\
	d\hat\rho_2=\frac{1}{2}\left[\sigma^z\cdot\hat\rho_2(d\omega_{21}-id\omega^*_{12})-
	\hat\rho_2\cdot\sigma^z(id\omega^*_{12}+d\omega_{21})\right].
	\end{split}
	\label{dynamics}
\end{equation}
One can easily prove that the solution of these two stochastic equations are
\begin{equation}
	 \hat\rho_1(t)=\left(
	 \begin{array}{cc}
    e^{i\omega^*_{21}}\rho_{1}^{11} & e^{\omega_{12}}\rho_{1}^{12} \\
    e^{-\omega_{12}}\rho_{1}^{21} & e^{-i \omega^*_{21}}\rho_{1}^{22}
	\end{array}
	\right). 
\end{equation}	
Here, we have introduced explicitly the elements of the initial state
$\rho_1^{ii}$. $\hat\rho_2(t)$ can be obtained from this expression by substituting
$1\rightarrow2$ in the subscript of the initial state, and by swapping
$1\leftrightarrow 2$ in the subscript of $\omega$. Some straightforward algebra
now leads to the expression of $\tilde \rho(t)$, and with the properties of the
averages we will discuss in a moment, to the final expression for the total
density matrix $\hat \rho(t)$. We can prove that the
density matrix obtained in this way is identical to the one obtained by
evaluating $\hat\rho(t)=\exp(-iHt)\hat\rho(0)\exp(iHt)$ where
$\hat\rho(0)=\hat\rho_1(0)\otimes\hat\rho_2(0)$. To obtain these results, we
need to evaluate the stochastic averages of two functions: namely, both $\exp(\omega_{12}+\omega_{21})$ and
$\exp(\omega_{12}+i\omega^*_{12})$. To do that, we observe that for any real
Wiener process, $p$, and any complex number, $\alpha$, we have
$\overline{\exp(\alpha p)}=\exp(-\alpha^2 t/2)$~\footnote{This can be easily
proven by considering the stochastic differential equation solved by
$\exp(\alpha p)$ according to Ito's calculus where $p$ is a real white noise, then  taking the average, and solving the differential equation for
$\overline{\exp(\alpha p)}$ obtained in this way.}. We easily then arrive at
$\overline{\exp(\omega_{12}+\omega_{21})}=1$ and
$\overline{\exp(\omega_{12}+i\omega^*_{12})}=\exp(-2it)$. We can now evaluate
the reduced density matrix for the spin 1, $\hat\rho_1^R(t)$. Again with some
straightforward algebra, and assuming that $\mathrm{tr}\hat\rho_1(0)=1$, we
obtain
\begin{equation}
	\begin{split}
	\hat\rho_1^R(t)
	=\left(
	 \begin{array}{cc}
    \rho^{11}_1 & \rho_1^{12}(t) \\
   \left(\rho_1^{12}(t)\right)^* & \rho_{1}^{22}
	\end{array}
	 \right)
	\end{split}
\end{equation}
where $\rho_1^{12}(t)=\rho^{12}_1\left(\cos
2t+i(\rho_{2}^{11}-\rho_{2}^{22})\sin 2t\right)$. With the reduced density
matrix we can therefore calculate the expectation value of any single spin
observable, e.g., $\langle\sigma_1^z\rangle=\rho_1^{11}-\rho_2^{22}$ and
$\langle\sigma^x_1\rangle=2\mathrm{Re}(\rho_1^{12})\cos(2t)-2\mathrm{Im}(\rho_1^
{12})(\rho_1^{22}-\rho_1^{11})\sin(2t)$.

The generalization to a chain of $N$ interacting spins 1/2 therefore follows in the
same footsteps. If the start with the Hamiltonian $\hat H=\sum_{i,j=1}^N
\lambda_{i,j}\sigma_i^z\otimes\sigma_j^z$, with $\lambda_{i,i}=0$ and use
Eq.~(\ref{effective_dynamics}), we obtain the single spin stochastic density
matrix
\begin{equation}
	\hat\rho_i(t)=\left(
	\begin{array}{cc}
		\rho_i^{11}e^{i\sum_{j=1}^N\sqrt{\lambda_{i,j}}\omega^*_{j,i}} & \rho_i^{12}e^{\sum_{j=1}^N\sqrt{\lambda_{i,j}}\omega_{i,j}}\\
		\rho_i^{21}e^{-\sum_{j=1}^N\sqrt{\lambda_{i,j}}\omega_{i,j}} & \rho_i^{22}e^{-i\sum_{j=1}^N\sqrt{\lambda_{i,j}}\omega^*_{j,i}} 
	\end{array}
	\right).
\end{equation}
From this result, deriving the single spin reduced density matrix is now straightforward,
\begin{equation}
	\begin{split}
	\hat\rho_i^R(t)=\overline{\prod_{\substack{k=1\\k\not =i}}^N \mathrm{tr}\hat\rho_k(t) \hat\rho_i(t)}
	=\left(
	\begin{array}{cc}
		\rho_i^{11} & f_i(t)\\
		f_i(t)^* & \rho_i^{22} 
	\end{array}
	\right),
	\end{split}
\end{equation}
where we have introduced the functions $f_i(t)=\rho_i^{12}\prod_{k=1,~k\not
=i}^N\left(\rho_k^{11}\exp(-2i\lambda_{ik}t)+\rho_k^{22}\exp(2i\lambda_{ik}t)\right)$. In the same way, we can build the 2-spin reduced density matrix starting
from its definition
$\hat\rho_{i,j}^R(t)=\overline{\prod_{\substack{k=1\\k\not =i,j}}^N \mathrm{tr}\hat\rho_k(t) \hat\rho_i(t)\otimes\hat\rho_j(t)}
$. The knowledge of $\hat\rho_{i,j}^R(t)$ allows us to calculate the correlation functions that have been recently measured \cite{Richerme2014,Senko2014}, in accordance to new theoretical results \cite{Foss-Feig2013,VanDerWorm2013}. The present theory provides an easy way to reproduce and generalize those results.

\begin{center}
	\begin{figure}[ht!]
		\includegraphics[width=8cm]{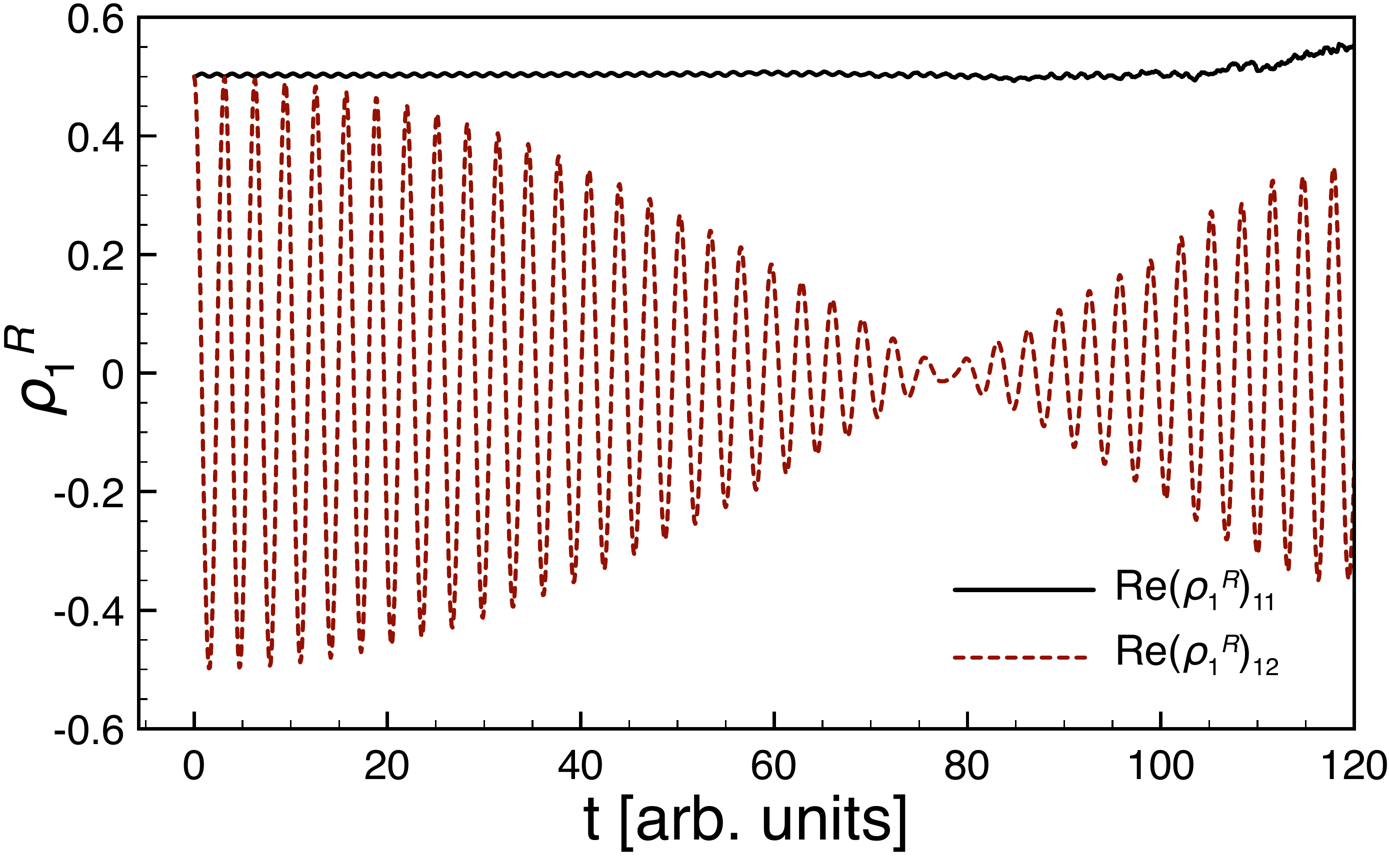}
		\caption{The time dynamics of the real
		parts of $(\rho_1^R)_{11}$ and $(\rho_1^R)_{21}$ calculated from
		Eqs.~(\ref{reduced_dynamics}) and (\ref{effective_dynamics}). As initial condition we have set the two spin in the ``mixed'' state. We have averaged
		over $10^6$ independent realizations of the white noises. For the calculation
		we have chosen $h_1=h_2=1$ and $\lambda=0.02$, and a time step, $\Delta
		t=0.01$. We can see that after $t\sim 100$ the finite number of realizations and the finite numerical accuracy in solving the stochastic differential equations are introducing an error in both $\rho_1^{11}$ and $\rho_1^{12}$.}
		\label{fig1}
	\end{figure}
\end{center}

We can investigate how the presence of a magnetic field, for example in the
$\hat x$ direction affects the dynamics. To do that, we consider two
interacting spins with the total Hamiltonian given by $\hat
H=h_1\sigma_1^x+h_2\sigma_2^x+\lambda\sigma_1^z\otimes\sigma_2^z$. We can write
down the stochastic equations of motion for the two single-spin density
matrices, but their analytic solution of little use since $\sigma_z$ and
$\sigma_x$ do not commute and we have to revert to a numerical solution of the equations of motion. 
In Fig.~\ref{fig1}, we report the dynamics of a few
elements of the density matrix $\hat\rho_1^R(t)$ calculated using
Eqs.~(\ref{reduced_dynamics}) and (\ref{effective_dynamics}), after taking the
average of $10^6$ independent realizations of the white noises. As initial condition we have assumed the two spins are in the mixed state, $\rho_1^{i,j}=\rho_2^{i,j}=1/2$ for any $i$ and $j$.

\begin{figure}[ht!]
	\includegraphics[width=8cm]{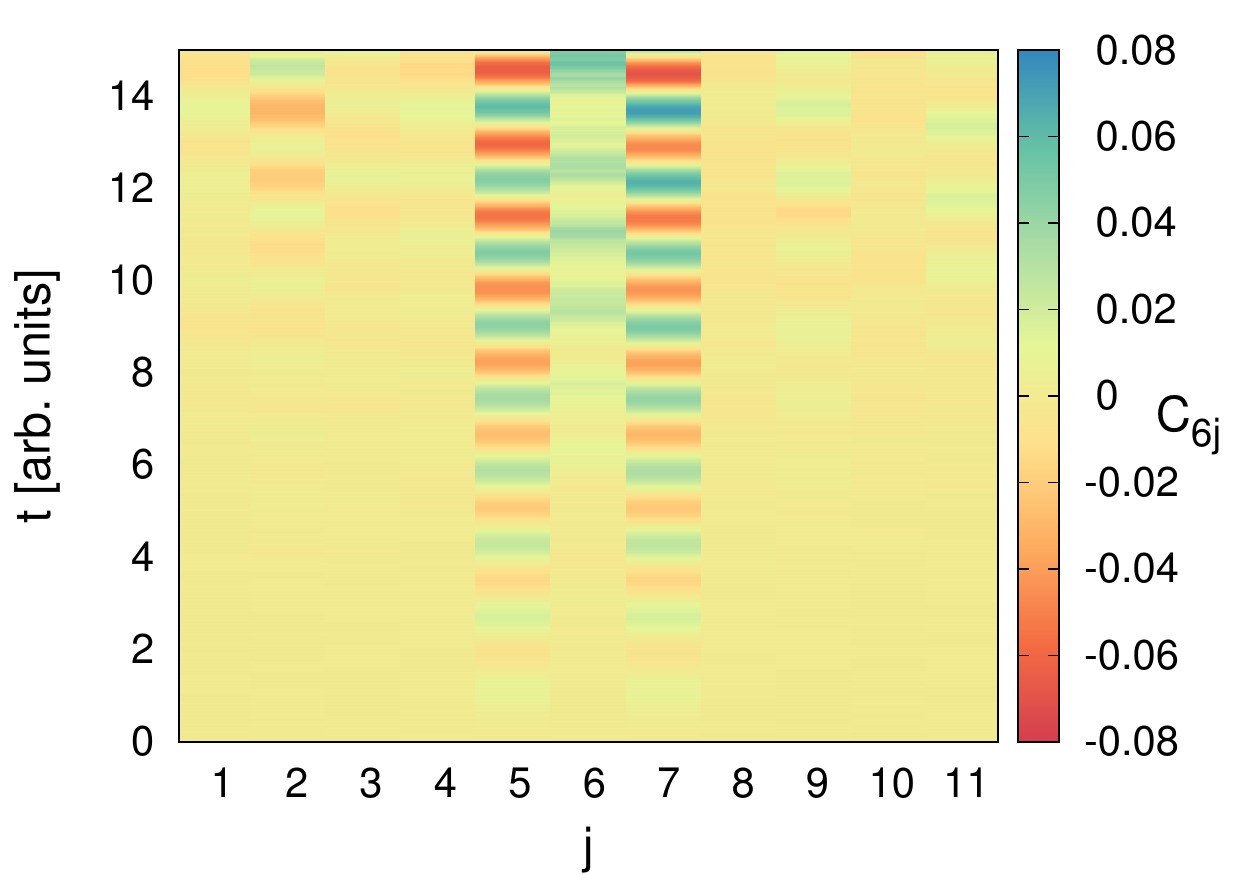}
	\caption{Dynamics of the correlation function $C_{6,j}(t)$ in time. We have chosen for the parameters $h=1$ and $\lambda=0.01$. All the spin but $6$ are in a mixed state, while spin $6$ starts in the state ``up''. We have chosen a time step of $\Delta t=0.002$ and averaged over 10000 realizations of the stochastic noise.}
	\label{fig3}
\end{figure}

The possibility to measure in time, and locally the time evolution of the
correlation between two spins has recently captured a lot of attention.
\cite{Richerme2014,Jurcevic2014} In the experimental set-up, we could change
the spin-spin interaction in such a way to explore the transition between a
XY-model to Ising model. In particular, this possibility has been investigated
in linear spin chains. Of particular interest is the spin-spin correlation
function, since it can be seen as a way to measure the spin propagation speed
and correlation time. To show how the present formalism can be applied to this case, we
consider the Hamiltonian $\hat H= \sum_{i=1}^N h_i
\sigma_i^z+1/2\sum_{i,j}^N\lambda_{ij}\left(\sigma_i^x\otimes\sigma_j^x\right)$,
 describing a chain of $N$ interacting spins in the presence of a static
magnetic field $\vec B_i=(0,0,h_i)$. In the following, we consider the case of
a chain of $11$ spins, in the presence of a uniform magnetic field, $h_i=h$ for
any $i$, and where the interaction if restricted to first neighbors. A
quantity of interest for these systems is the time evolution of the correlation
function, defined as $C_{i,j}=\langle \sigma^z_i\sigma^z_j\rangle-\langle
\sigma^z_i\rangle\langle\sigma^z_j\rangle$. In Fig. \ref{fig3} we report the
time evolution of $C_{6,j}(t)$. We have chosen as initial condition that all
the spin but $6$ are in a mixed state, while at $t=0$ the state of spin $6$ is
up, that is $\rho_6^{1,1}=1$, $\rho_6^{1,2}=\rho_6^{2,1}=\rho_6^{2,2}=0$, while
we have used $h=1$ and $\lambda_{ij}=0.01$ if $|i-j|=1$, or $0$ otherwise. We
have chosen a time step of $\Delta t=0.002$ and averaged over 10000
realizations of the stochastic noise. It is seen that the correlation grows
rapidly for the spin $5$ and $7$, while for the other spins, that are not
directly connected with spin $6$, it remains rather small. This can be compared with the results of the experiments reported in \cite{Richerme2014,Jurcevic2014}.

It is well known that the von Neumann equation is equivalent to the
Schr\"odinger equation to almost any purpose in standard quantum mechanics
\cite{Sakurai,Gardiner2000}. It then appears natural to extend the result of this Letter to
the many-body problem described by the time evolution of the many-body
wave-function, $|\Psi(t)\rangle$. We can prove, following essentially in the
footstep of the proof given for the many-body density matrix, than it is
possible to find stochastic single-particle wave-functions $|\psi_i\rangle$
such that, if at the initial time $|\Psi(0)\rangle=\bigotimes_{i=1}^N
|\psi_i(0)\rangle$, then for any subsequent time
$|\Psi(t)\rangle=\overline{\bigotimes_{i=1}^N |\psi_i(t)\rangle}$. The states
$\psi_i(t)$ evolve according to a stochastic Schr\"odinger equation
\cite{Strunz2000,Gardiner2000,Biele2012,DAgosta2008a,DAgosta2013a} which has a
form similar to Eq.~(\ref{effective_dynamics}). Interestingly, while the
standard relation $\rho(t)=|\Psi(t)\rangle\langle \Psi(t)|$ is satisfied, there
not exists a similar relation between the single-particles, i.e., in general we
should expect that $\rho_i(t)\not = |\psi_i(t)\rangle\langle\psi_i(t)|$.
Indeed, one can prove that $|\psi_i(t)\rangle\langle\psi_i(t)|$ has an
equation of motion that does not reduce to Eq.~(\ref{effective_dynamics}).
Finally, we would like to comment that a formalism based on the wave-function
$|\psi_i\rangle$ seems limited, since it is not clear how one could possibly
obtain information on the single-particle properties of the real many-body
problem. In fact, if we start with the definition of $\hat\rho_i^R(t)$ from
Eq.~(\ref{reduced_dynamics}), and use the wave-function $|\Psi(t)\rangle$
instead we get,
\begin{equation}
	\begin{split}
	\hat\rho_j^R(t) & =\mathrm{tr}\hat\rho(t)|_{\mathrm{but}~ j}=\mathrm{tr}\left(|\Psi(t)\rangle\langle\Psi(t)|\right)|_{\mathrm{but}~j}\\
	&=\mathrm{tr}\left(\overline{\bigotimes_{i=1}^N |\psi_i(t)\rangle}~\overline{\bigotimes_{i=1}^N \langle\psi_i(t)|}\right)_{\mathrm{but}~j}.
	\end{split}
\end{equation}
It is therefore not possible to swap the trace operation with the average over
the realizations of the stochastic noise, therefore precluding an alternative
way to Eq.~(\ref{reduced_dynamics}). Similar problems arise when one considers
any $n$-body reduced density matrix as obtained from the single-particle stochastic wave-functions.

In conclusion, we have shown that the dynamics of a many-body system, with
multiplicative two-body particle-particle interaction can be reduced to the
investigation of the dynamics of $N$ single particle stochastic systems. We
have shown how to calculate any reduced $n$-body properties
starting from the solution of these stochastic dynamical equations. We have
applied this formalism to the case of spin chains in the presence of
a finite magnetic field. 
%The investigation of more complex systems like for
%example the Hubbard model is currently under-way.

\section*{Methods}
Here we present the proof of Eq.~(\ref{exactrho}), 
based on the It\^o calculus \cite{Higham2001,Gardiner2000}.

{\bf Proof.} The proof of Eq.~(\ref{exactrho}), and that $\hat\rho(t)$ follows
the dynamics induced by the von Neumann's equation (\ref{von_neumann}) is a
straightforward generalization of the results of Shao on the dynamics of a
quantum system in contact with an external environment
\cite{Shao2004,Shao2010}. Let us consider the stochastic total density matrix,
$\tilde\rho(t)=\bigotimes_i\rho_i(t)$. According to It\^o's stochastic
calculus, the dynamics of $\tilde\rho$ is determined by,
$
d\tilde\rho(t)=\sum_{j=1}^N\rho_{1,j}\otimes d\rho_j\otimes\rho_{j+1,N+1}
+\sum_{k<j=1}^N\left(\rho_{1,k}\otimes d\rho_k\otimes \rho_{k+1,j}\otimes d\rho_j \otimes \rho_{j+1,N+1}\right),
$
where we have introduced the short-hand notation 
$
	\rho_{i,j}=\bigotimes_{k=i}^{j-1}\rho_k.
$
The proof that $\overline{\tilde\rho(t)}=\rho(t)$ then continues by using
Eq.~(\ref{effective_dynamics}), Eq.~(\ref{white_noise}), and $[\hat x_i,\hat
\rho_j]=0$ if $i\not = j$, to arrive at
$
d\overline{\tilde\rho(t)}=-i\left[\hat H,\overline{\tilde\rho(t)}\right]dt.
$
Due to the linearity of this equation of motion, and that by definition
$\overline{\tilde\rho(0)}=\hat\rho(0)$, we conclude that
$\overline{\tilde\rho(t)}=\hat\rho(t)$ at any time $t\ge0$.$\square$

%\bibliographystyle{naturemag}
%\bibliography{library.bib}


\end{document}